# On the operation of a Micropattern Gaseous UV-Photomultiplier in Liquid-Xenon

**S. Duval**[a*]**, A. Breskin**[b]**, R. Budnik**[b]**, W.T. Chen**[a]**, H. Carduner**[a]**, M. Cortesi**[b]**, J.P. Cussonneau**[a]**, J. Donnard**[a]**, J. Lamblin**[a]**, P. Le Ray**[a]**, E. Morteau**[a]**, T. Oger**[a]**, J.S. Stutzmann**[a] **and D. Thers**[a]**.**

[a] *Subatech, Ecole des Mines, CNRS/IN2P3 and Université de Nantes,*
  *44307 Nantes, France*

[b] *Department of Astrophysics and Particle Physics, Weizmann Institute of Science,*
  *76100 Rehovot, Israel*
  *E-mail*: samuel.duval@subatech.in2p3.fr

ABSTRACT: Operation results are presented of a UV-sensitive gaseous photomultiplier (GPM) coupled through a $MgF_2$ window to a liquid-xenon scintillator. It consisted of a reflective CsI photocathode deposited on top of a THick Gaseous Electron Multiplier (THGEM); further multiplication stages were either a second THGEM or a Parallel Ionization Multiplier (PIM) followed by a MICROMEsh GAseous Structure (MICROMEGAS). The GPM operated in gas-flow mode with non-condensable gas mixtures. Gains of $10^4$ were measured with a CsI-coated double-THGEM detector in $Ne/CH_4$ (95:5), $Ne/CF_4$ (95:5) and $Ne/CH_4/CF_4$ (90:5:5), with soft X-rays at 173 K. Scintillation signals induced by alpha particles in liquid xenon were measured here for the first time with a double-THGEM GPM in $He/CH_4$ (92.5:7.5) and a triple-structure THGEM/PIM/MICROMEGAS GPM in $Ne/CH_4$ (90:10) with a fast-current preamplifier.

---

[*] Corresponding author

# Contents

## 1. Introduction

We present here first results on the detection of radiation-induced primary-scintillation in liquid xenon with a gaseous photomultiplier coupled to the liquid.

Over the past decade there has been an increasing interest in noble-liquid detectors, mainly focused towards Dark Matter search [1]; these are single-phase scintillation and ionization detectors and more complex dual-phase (liquid/gas) devices [2]. Other applications using mostly liquid xenon (LXe) detection medium are Gamma detectors for PET [3], a Compton Camera for applications in astrophysics [4] and the recently proposed single-phase LXe Compton Telescope for "3γ imaging" in combination with PET [5].

While the radiation-induced primary- and secondary-scintillation light in noble-liquid detectors is detected with dedicated costly vacuum photomultipliers [1], there have been attempts to develop gaseous photomultipliers (GPM [6]) capable of operation at cryogenic temperatures and of large-area coverage [7][8][9]. Room-temperature GPMs with CsI photocathodes coupled to wire chambers have been proposed and investigated for recording Xe scintillation light [10][11]. Our "3γ imaging" detection concept consists of a 3D localization of a ($\beta^+$, γ) emitter, $^{44}$Sc, by intersecting the line-of-response (LOR) annihilation gamma rays (in a PET) with the direction of the third gamma ray (of 1,157 MeV) measured with a LXe Compton Telescope [5]. The temporal and spatial coincidence of the three photons provides a unique event-by-event localization of the radioisotope along the LOR. However, this technique requires a large acceptance (sensitive volume) where Compton scattering is dominant around 1 MeV. This can be accomplished with the implementation of a LXe time-projection chamber (TPC), as demonstrated by the LXeGRIT experiment [4]. The interaction of radiation with LXe induces a fast UV scintillation-light flash used for the event time tagging and interaction-depth information; a simultaneous ionization charge induced and collected in the liquid provides 2D position and energy. According to Compton kinematics, the measurement of the position and energy of the two first interactions is sufficient for the reconstruction of the third gamma ray direction [12].

In order to improve the Compton Telescope trigger efficiency the Subatech and Weizmann Institute teams proposed to maximize the total detection area, while keeping a homogenous LXe volume, by substituting the traditional array of vacuum-photomultipliers with a single large-area cryogenic gaseous photomultiplier [13]; it is separated from the liquid by a flat UV-transparent window.

We present here results of measurements performed with a 32 mm diameter (active area) GPM prototype operated at cryogenic conditions in $Ne/CH_4$ and $Ne/CF_4$. Various detector configurations have been investigated, consisting of a cascaded electron multiplier with different elements: THick Gaseous Electron Multipliers (THGEM [6]), Parallel Ionization Multipliers (PIM [14]) and MICROMEsh GAseous Structure (MICROMEGAS [15]); the first multiplier (THGEM) was coated with a reflective CsI photocathode. Gain curves measured at temperatures down to 173 K are presented. Signals of $^{238}$Pu alpha-particles interacting with LXe were recorded for the first time with a GPM coupled to LXe; they are compared with that of a vacuum PMT.

## 2. Materials and Methods

### 2.1 Experimental setup

The general layout of the cryogenic GPM prototype is schematically shown in figure 1. The envelope of the detector was made of DN 250 and 160 CF flanges in 304L stainless steel; a $MgF_2$ entrance window (thickness = 2 mm, diameter = 23 mm, LewVac) was sealed on a DN40 CF flange with a Kovar® alloy bellow. Standard high-voltage and signal feedthroughs were used to ensure tightness to keep a vacuum of $10^{-9}$ bar inside the insulated layer of the cryostat and 1100 mbar of counting-gas within the detector vessel.

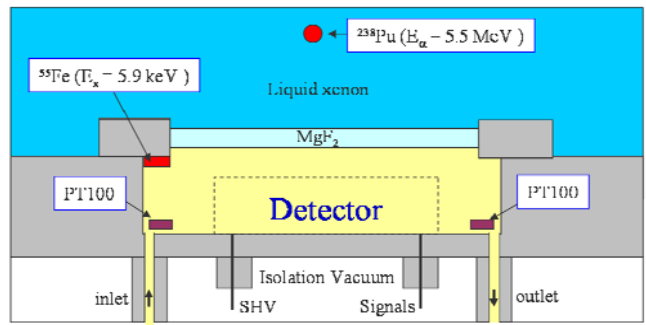

**Figure 1**. Schematic layout of the gaseous photomultiplier.

The GPM was cooled down and immersed within LXe for about 24 hours in order to obtain a stabilized gas-mixture temperature of 173 K. The cryogenic LXe system (described in detail in [16][17]) and the gas manifold of the GPM are schematically depicted in figure 2. The photon detector was decoupled from LXe by the $MgF_2$ optical window. Non-condensable Ne and He mixtures with $CH_4$ and $CF_4$ quenchers were flushed through the GPM with gas-flow controllers (EL-FLOW Bronkhorst). The flow did not exceed 2 l/h (at normal conditions) to assure the gas-temperature homogeneity between the inlet and outlet, necessary to maintain gain stability. A temperature difference of 1 K was monitored by two PT100 temperature sensors located at the gas inlet and outlet. The pressure within the GPM was regulated by the gas-flow controller to 1100 mbar, to minimize mechanical stresses on the $MgF_2$ window; the gas was evacuated through an oil bubbler to avoid air reflux.

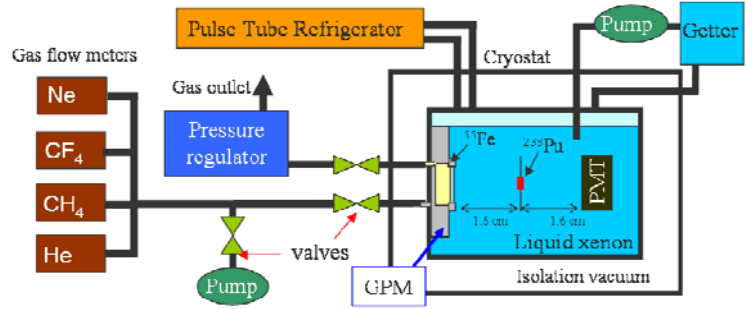

**Figure 2**. Schematic layout of the cryogenic system and the gas manifold of the gaseous photomultiplier (GPM); the latter was immersed within liquid xenon.

The xenon was liquefied by a remote pulse-tube refrigerator (Model PC150 (S/N: MP8000B-46)), which delivered 100 watts of cooling power at 165 K; the pressure was kept at 1345 mbar. Xe was extracted from the liquid phase and continuously purified in gas phase, to eliminate impurities that could affect the luminescence yield. This was done by circulating throughout a micro-pump (ENOMOTO Model MX-808ST-S) and a getter (SAES MonoTorr® R Phase II Purifier PS4-MT3/15-R/N-1/2) at a flow rate of 5 $l_{(gas)}$/min.

As mentioned above, the GPM investigated in this work consisted of a THGEM coated with a reflective CsI photocathode [18][19], followed by a second THGEM ("double-THGEM") or by two other electron multipliers (defined here as "triple-structure detector").

The "double-THGEM" (figure 3a) configuration consisted of two THGEMs in cascade, the latter are made of 400 μm thick double-sided Cu/Au Printed Circuit Board (PCB) with an array of mechanically drilled holes, of 300 μm in diameter and 700 μm apart [6]. 50 μm rims etched around the holes considerably reduced discharge probability. The THGEM-hole geometry permitted depositing reflective CsI photocathode on top of the electrode, with high surface coverage leading to large effective QE values [18]. The operational voltages allowed reaching high electric field strengths at the photocathode surface, required to reduce photoelectron backscattering, with a resulting good extraction efficiency from CsI and close to their full collection efficiency into the THEM holes [18][20]. A reflective CsI photocathode of 200 nm thickness was thermally-deposited on the top surface of the first THGEM at a deposition rate of ~ 10 Å.s$^{-1}$ in a vacuum evaporator evacuated to $10^{-6}$ bars. It was stored under nitrogen prior to transfer in air to the detector. A cathode grid was placed 4 mm above the first THGEM to permit the application of a drift field ($E_{drift}$) adjusted for gain measurements (with X-rays) or UV scintillation-photon recording - as described in the following sections 2.2 and 2.3.

The "triple-structure detector" (figure 3b) consisted of a CsI-coated THGEM (described above), followed by two micromesh structures known as PIM and MICROMEGAS; they acted as additional electron multipliers as well as efficient avalanche-ion blockers [21]. This detector configuration was investigated in view of its potential capability of considerably reducing the photocathode aging by ion bombardment [22]. The MICROMEGAS consisted of a 5 μm thick copper electrode with 30 μm diameter holes spaced by 60 μm. The mesh was maintained at 50 μm from the anode by Kapton pillars. The PIM consisted of two 5 μm thick electroformed nickel grids with different mesh parameters (50.8 μm and 37.9 μm) with beams width of 11.7 μm, separated by a Kapton spacer defining an amplification gap of 125 μm. The spacer was a Kapton laser-cut grid with square openings of 3 mm side and 50 μm wide beams.

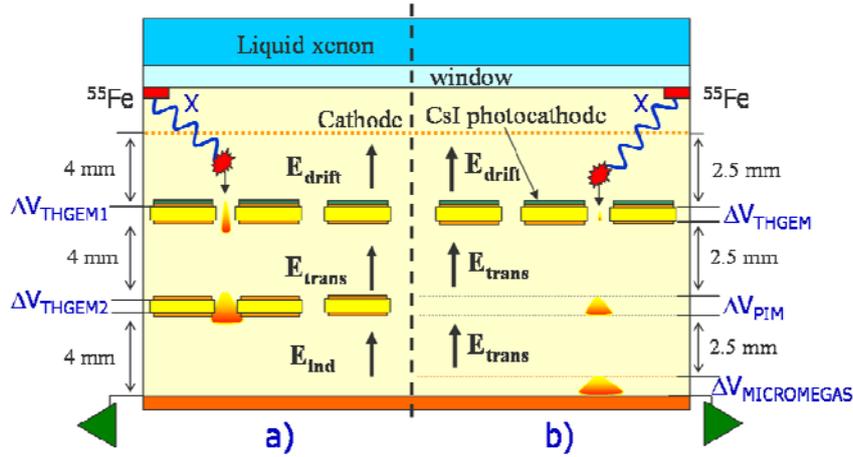

**Figure 3**. Schematic drawing of the double-THGEM detector (a) and in the triple-structure (b) in a pulse-mode gain measurement configuration, with a 5.9 keV $^{55}$Fe X-ray source.

## 2.2 Gain measurements in pulse-mode with a $^{55}$Fe source

Gain characterizations of the GPM coupled to LXe were done with 5.9 KeV X-rays from a $^{55}$Fe source glued inside the GPM (Figure 3), inducing ~200) primary electrons per interaction in the Ne mixtures (for $W_{Ne}$ = 36 eV). A drift field ($E_{drift}$) was applied between a nickel cathode grid (70 lpi) placed above the first THGEM, to collect the charges into the THGEM holes.

In the double-THGEM detector configuration the amplification took place in the holes of both THGEMs; charges were transferred from the first multiplier to the second one, upon application of a suitable transfer field ($E_{trans}$) between the two multiplication stages. In the triple-structure detector, after amplification in the THGEM holes, the resulting avalanche electrons were further transferred (under $E_{trans}$) and multiplied in the PIM and than in the MICROMEGAS elements.

The dimensions of the different gaps are indicated in figure 3. The charge motion in the last, (induction), gap of each detector configuration induced pulses into the respective anodes (Figure 3). They were measured with an electronic chain located outside the cryostat at room temperature: an ORTEC 142 IH charge-sensitive preamplifier followed by an ORTEC 655 Dual amplifier (shaping time = 500 ns), a quad linear fan-in fan-out NIM-N625, a six-channel discriminator NIM MODEL 711 Phillips Scientific and a dual timer MOD 2255B CAEN used to trigger the acquisition of an ADC peak sensing Mod V785N. The electronic chain was calibrated by injecting a known charge through the 1 pF preamplifier input test capacitance with an Agilent 33250A pulser.

## 2.3 Measurement of [238]Pu induced scintillation pulses in liquid-xenon

A [238]Pu point source emitting 5.5 MeV alpha particles electrochemically deposed on a platinum wire of 700 μm diameter was placed inside liquid-xenon at 1.6 cm from the $MgF_2$ window and 4.3 cm from the cathode grid. A PMT (Hamamatsu R7600 06 MOD-ASSY) facing the GPM was placed at 1.6 cm from the source to trigger the GPM acquisition on alpha-induced scintillation signals as shown in figure 2. The cathode and photocathode were kept at the same potential ($E_{drift} = 0$) in order to efficiently focus the extracted photoelectrons into THGEM holes [18] (see figure 4). The rest of the remaining detector setup was the same as described in section 2.2. A fast home-made current preamplifier based on AD8015 wideband/differential output trans-impedance preamplifier, was situated outside the cryostat; it was used to visualize the current induced by the electron motion in the induction gaps. Signals were recorded by a LECROY LT374 M oscilloscope.

Assuming a mean energy value per electron/ion pair $W_{ph}(\alpha) = 17.9$ eV [23], the expected mean number of photoelectrons extracted from the CsI photocathode per alpha-particle stopped in the liquid was estimated (by a Monte-Carlo simulation) to be around 650. The following parameters were taken into account: the scintillation-light attenuation length (1 meter [24]), the refractive index of xenon ($n_{LXe} = 1.61$ [3]), of $MgF_2$ ($n_{MgF2} = 1.44$ [25]) and on the Ne gas mixture ($n_{gas} = 1.00$) at 178 nm, the mean alpha-particle interaction length of 40 μm in LXe [26], neglecting the Rayleigh scattering. The effective QE of reflective CsI photocathode deposed on the THGEM electrode was assumed to be 22 % according to [18][20].

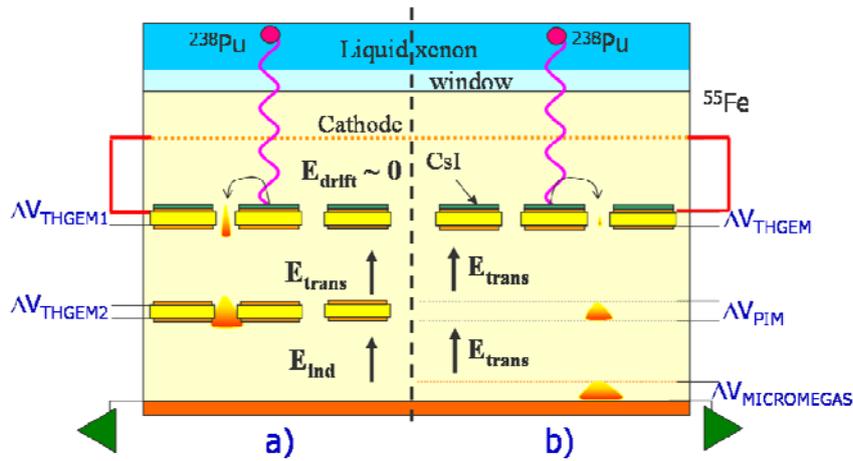

**Figure 4**. Schematic drawing of the double-THGEM detector (a) and triple-structure detector (b) in the scintillation measurement configuration, with a 5.5 MeV [238]Pu alpha source.

## 3. Results

### 3.1 Gain measurements with X-rays in pulse mode

Figure 5 shows gain curves recorded with the double-THGEM detector at room temperature and at 173 K in 1100 mbar neon mixtures Ne/$CH_4$ (95:5), Ne/$CF_4$ (95:5) and Ne/$CH_4$/$CF_4$ (90:5:5); the gain was measured with a low-rate (few Hz) 5.9 keV X-ray source. At cryogenic temperature the GPM was coupled to LXe (Figure 3). The gain maxima correspond to THGEM sparking limits in the different gas mixtures. Error bars are associated to the measured widths of the X-ray peaks (RMS). Large error bars correspond to broad or unresolved X-ray distributions. Gain curves recorded in Ne/$CF_4$ (95:5) at room temperature showed some saturation at high THGEM polarization voltages; the interpretation of this effect, not observed in our previous sets of measurements in another system [16] could originate from some gas impurities; its interpretation requires additional studies. In both mixtures (Ne/$CH_4$ and Ne/$CF_4$) gains of about $10^5$ were reached at room temperature, with slightly larger applied voltages with $CF_4$ for the same total gain. Similar trend was observed at low temperature (T = 173 K) but the maximum reachable gains were about ten-fold lower – in agreement with recent results in [27]. An increase of the gas quencher concentration by mixing 5%$CH_4$ and 5%$CF_4$ with Ne, (full inverse triangles) resulted in a further lower maximal detector gain, reached at higher voltages. The slight difference between the gain curves in Ne/5%CH4 and Ne/5%CF4 measured here and in previous works [18][19] could be due to small variations in the gas purities and mixtures in the different experimental systems (at different laboratories).

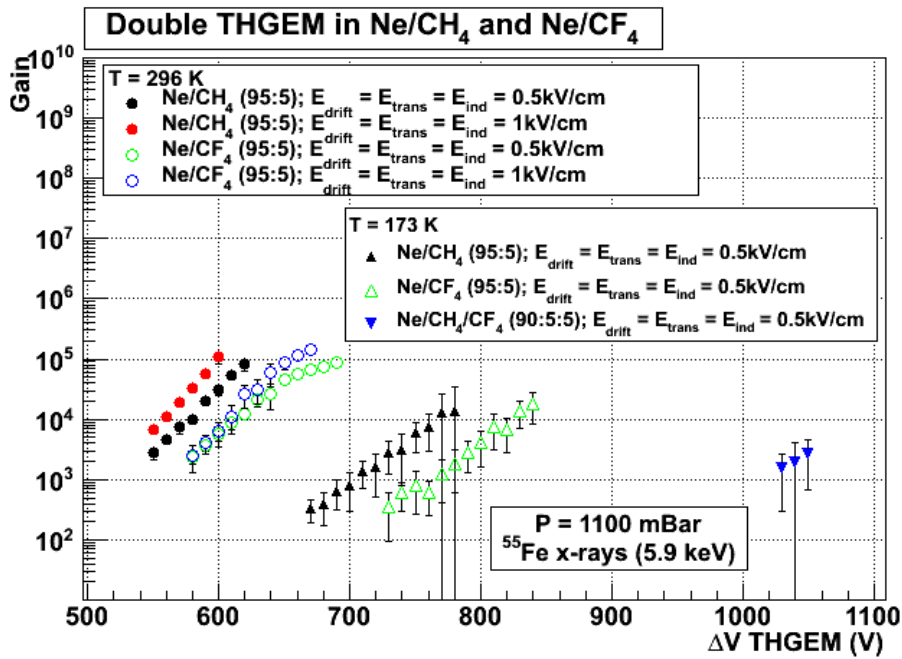

**Figure 5**. Charge-gain curves obtained with the double-THGEM detector operated with 5.9 keV X-rays at 296 K (circles) and 173 K (triangles) in 1100 mbar of neon mixtures for different drift-, transfer- and induction-field configurations (indicated in the figure 3).

The gain of the triple-structure detector THGEM/PIM/MICROMEGAS was measured at room temperature in Ne/$CH_4$ (95:5) and Ne/$CH_4$ (90:10) mixtures as shown in figure 6. One can notice that gains of about $10^5$ were also reached with this detector despite a small amplification factor (~ 100) in the THGEM element (estimated from the applied $\Delta V$ = 500 V and 625 V in Ne with 5% and 10 % of methane [18]). However, gain instabilities were encountered at low temperature and did not permit precise gain measurements; the pulses amplitude was comparable to those observed with the double-THGEM.

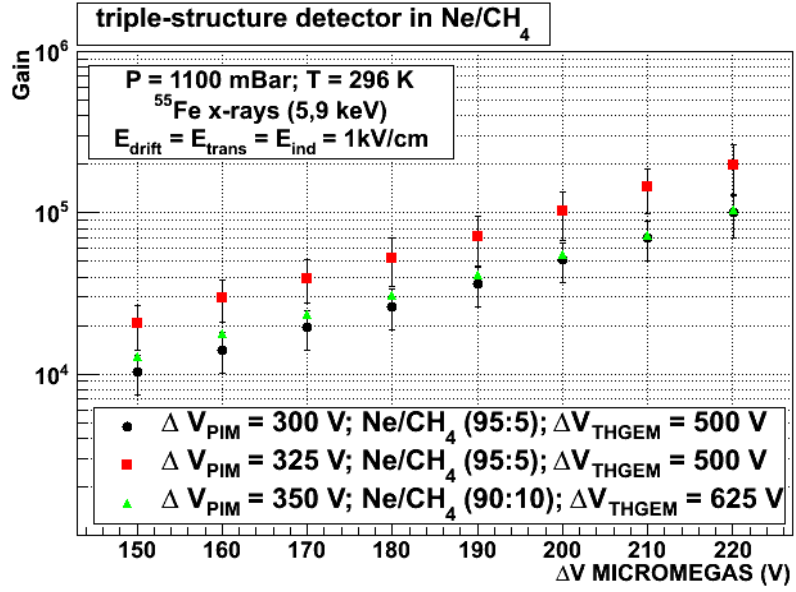

**Figure 6**. Charge-gain curves obtained with the triple-structure detector with 5.9 keV X-rays at 296 K in Ne/CH$_4$ (95:5) and Ne/CH$_4$ (90:10) for different THGEM and PIM polarization voltages.

### 3.2 Observation of $^{238}$Pu induced scintillation pulses in liquid-xenon

Scintillation pulses induced by $^{238}$Pu alpha-particle interactions with liquid xenon (Figure 4) were observed at 173 K with the double-THGEM detector coupled to a fast current preamplifier in 1100 mbar of He/CH$_4$ (92.5:7.5) as shown in figure 7. Helium was used here instead of Ne to achieve higher gains in order to visualize pulses with our fast-current preamplifier. Indeed, with the latter, the amplitude of scintillation-induced pulses in Ne/CH$_4$ (95:5) was at the limit of our electronics noise. We can see the same event recorded by the PMT in channel 1 and the GPM in channel 4 in coincidence. THGEM differential potentials were: $\Delta V_{THGEM1} = 1040$ V, $\Delta V_{THGEM2} = 1080$ V; the other electric field settings were: $E_{drift} = 0$ kV/cm, $E_{trans} = 0.5$ kV/cm and $E_{ind} = 1.25$ kV/cm. The PMT pulse reproduces the well-known liquid-xenon scintillation decay shape [28]. The GPM's pulse shape is derived from a convolution of the LXe scintillation photons arrival time, the avalanche-charge motion within the 4 mm induction gap, and the preamplifier's response. It rises in ~ 200 ns to reach a plateau of 120 ns; its decay corresponds to the beginning of charge collection on the anode.

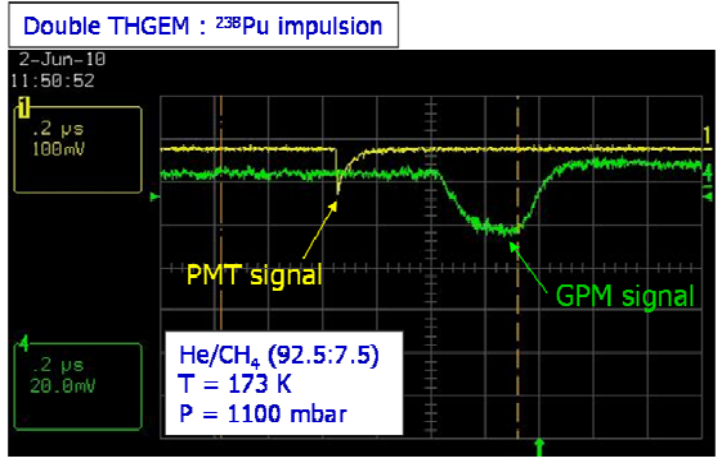

**Figure 7**. Scintillation signals induced by $^{238}$Pu alpha source in liquid-xenon recorded in coincidence, recorded in the GPM (double-THGEM) and the PMT. The GPM signal was read with a fast-current preamplifier. (Gas mixture: He/CH$_4$ (92.5:7.5), T = 173 K and P = 1100 mbar).

Figure 8 shows for comparison the X-ray induced pulses recorded in the same conditions and with same electronics. Only the drift field was modified to collect the primary electrons. The THGEM differential potentials were: $\Delta V_{THGEM1}$ = 1040 V, $\Delta V_{THGEM2}$ = 1080 V and the other electric fields were: $E_{drift}$ = 0.5 kV/cm, $E_{trans}$ = 0.5 kV/cm and $E_{ind}$ = 1.25 kV/cm. Corresponding gain measurements were not possible because of the unresolved $^{55}$Fe spectra recorded in the present helium mixture. However, the average signal's amplitude appears to be larger than that recorded with the $^{238}$Pu source (Figure 7).

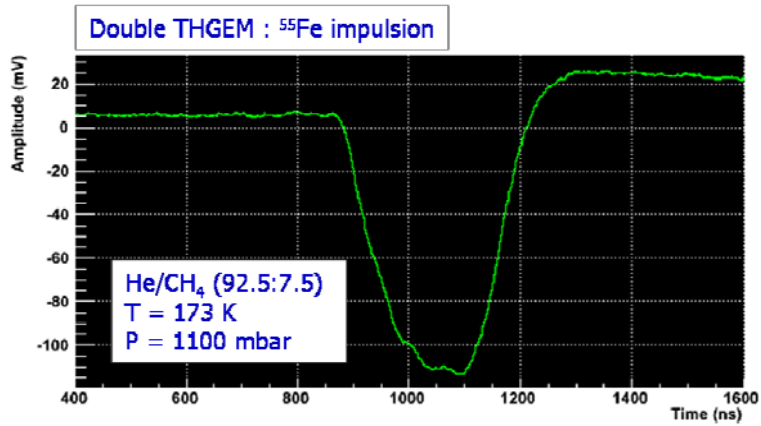

**Figure 8**. Gaseous photomultiplier (double-THGEM) signal induced by $^{55}$Fe X-rays, recorded with a fast-current preamplifier. (Gas mixture: He/CH$_4$ (92.5:7.5), T = 173 K, P = 1100 mbar).

Scintillation signals were also observed with the triple-structure detector, operated (unlike the double-THGEM) in Ne/CH$_4$. Figure 9 shows a pulse recorded in 1100 mbar of Ne/CH$_4$ (90:10) at T = 173 K. The differential potentials were: $\Delta V_{THGEM}$ = 700 V, $\Delta V_{PIM}$ = 625 V and $\Delta V_{MICROMEGAS}$ = 265 V. The preamplifier signal was filtered (20 MHz) to suppress high-frequency noise induced by the high voltage power supply; it slightly smoothened the pulse shape. The 3-4 fold larger pulse height recorded in the Ne-mixture and triple-structure (Figure 9 compared to Figure 8) could be due to the combination of somewhat higher gain, different pulse-shape (due to different multiplication and pulse-formation mechanisms) and larger effective QE of CsI due to reduced backscattering in Ne mixture. As expected, the scintillation signal appeared to rise faster than in the double-THGEM detector; it is due mainly to the very narrow, 50 μm, amplification and induction gap of the MICROMEGAS (compared to a 4 mm induction gap used here for the double-THGEM); the difference in drift velocity between Ne- and He-mixtures plays here only a minor effect. The signal decay seems to reproduce the liquid-scintillation decay but oscillations due to the readout electronics blurred the decaying pulse shape.

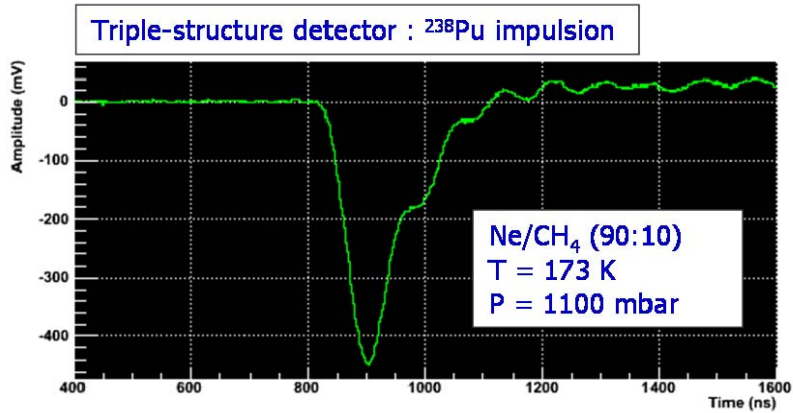

**Figure 9**. Gaseous photomultiplier (triple-structure) scintillation signal induced by $^{238}$Pu alpha-particles in liquid-xenon recorded with a fast-current preamplifier. (Gas mixture: Ne/CH$_4$ (90:10), T = 173 K, P = 1100 mbar).

## 4. Discussions & Conclusions

This work provided results on the operation of a GPM for recording primary radiation-induced scintillation light in liquid xenon. The GPM was either a double-THGEM or a triple-structure THGEM/PIM/MICROMEGAS - both with a reflective CsI photocathode deposited on the top THGEM face. Gains of about $10^4$ where reached at cryogenic conditions (173 K) with 5.9 KeV X-rays, with a double-THGEM detector operated in Ne/CH$_4$ (95:5) and Ne/CF$_4$ (95:5) mixtures. Gains of the same order were reached in the triple-structure detector: THGEM/PIM/MICROMEGAS; however in this structure we noticed some gain variations and electrical instabilities. The latter, that prevented precise multiplication measurements, could be due to the higher electric fields required by the increased gas density at low temperatures. Furthermore, we suspect some deformation of the metal-grid frames at cryogenic conditions.

Nevertheless, to the best of our knowledge, for the first time scintillation signals induced by alpha-particles were recorded with a gaseous photomultiplier directly in contact with liquid xenon - using micropattern electron multipliers coated with a reflective CsI photocathode. The amplitudes of the scintillation signals observed with the GPM were surprisingly small in comparison with 5.9 keV X-ray induced pulses recorded in the same conditions and electronics (see section 2.3). Even if, according to [29] photoelectron extraction from CsI into He mixtures is only about 60 % of the one in Ne/5%CH$_4$ for electric field of about 1-2 kV/cm [18][30], He/CH$_4$ mixture (92.5:7.5) was used here to reach a higher total detector gain (in double-THGEM) required for recording the scintillation pulses with the fast current preamplifier. Ne/10%CH$_4$ and Ne/10%CF$_4$ mixtures were used with the triple-structure detector to reach higher photoelectron extraction efficiency [18][30]. The fact that the alpha-particle scintillation induced in LXe resulted in a rather low yield of detected photoelectrons (e.g. compared to 5.9 keV X-rays) could indicate a low QE of the CsI photocathode used in our experiments. We suppose that a deterioration of the photocathode could have occurred during its transport (from Rehovot to Nantes, though under N$_2$), the detector assembly or during the cooling to LXe temperatures. The cooling process could result in a condensation of impurities on the photocathode surface, affecting the photoemission. This was recently shown in [27], where the relative QE of CsI dropped at low temperatures and recovered after warm up in Ne/5%CH$_4$; in Ne/5%CF$_4$ it lost however permanently part of its emission properties, possibly due to water-vapor interaction with the surface in presence of CF$_4$. Another factor which might have affected the effective CsI QE could be the photoelectron backscattering effect enhanced by the increased gas density at low temperature. In the present conditions (T = 173 K and constant pressure of 1100 mbar), the gas density was 2.7 times higher than at room temperature. However, according to electron backscattering simulations [30] and to recent measurements [27] the gas-density effect is negligible and should not affect dramatically the effective CsI QE value. Some complementary systematic investigations are presently carried out in a high-purity setup to elucidate this effect. Other studies are directed towards the choice of the optimal application-tailored cascaded-multiplier configuration for low-temperature operation. It should take into account the requested detector gain, rapidity of response, radiation flux (e.g. affecting photocathode aging), large-area feasibility, readout electronics and cost.

**In summary**, we have made a first proof of feasibility of scintillation-photons recording from LXe with a gaseous photomultiplier immersed within the liquid and operating in a different counting gas. Such photon detectors, that have the potential of economically cover large area, could be of interest for various applications with different noble liquids; among them are large-volume detectors of rare events (e.g. Dark Matter, Neutrino Physics, Double Beta Decay, etc) as well as gamma detectors for medical imaging, particle physics and safety inspection.

### Acknowledgements

This work was partly supported by the region of Pays de la Loire, France, by the Israel Science foundation (Grant Number 477/10) and by the MINERVA Foundation with funding from the federal German Ministry for Education and Research (Grant number 710827). A. Breskin is the W.P. Reuther Professor of Research in the peaceful use of Atomic Energy.